\documentclass[12pt]{article}
\usepackage{amsfonts}
\usepackage{amssymb,amsmath,amsthm,latexsym}
\usepackage[numbers,sort&compress]{natbib}
\usepackage{amsmath}
\usepackage{indentfirst}
\allowdisplaybreaks[4]
\newtheorem{theorem}{Theorem}[section]

\newtheorem{cor}[theorem]{Corollary}
\newtheorem{lemma}[theorem]{Lemma}

\newtheorem{remark}[theorem]{Remark}
\newtheorem{define}[theorem]{Definition}
\newtheorem{example}[theorem]{Example}

\newcommand{\pf} { {\rm \noindent{\bf Proof.}} }

\numberwithin{equation}{section}
\begin{document}

\title{On new quantum codes from matrix product codes }
\author{
 Xiusheng Liu$^a${\thanks{Corresponding author. \newline  Email addresses: lxs6682@163.com(Xiusheng Liu), hdinh@kent.edu(H.Q. Dinh), hwlulu@aliyun.com(Hualu Liu), longyuhbpu@163.com(Long Yu)} }, Hai Q. Dinh$^b$, Hualu Liu$^a$, Long Yu$^a$}
\date{a. School of Mathematics and Physics, Hubei Polytechnic University, Huangshi 435003, China\\
      b. Department of Mathematical Sciences, Kent State University, 4314 Mahoning Avenue, Warren, OH 44483, USA }

\maketitle

%\vspace*{3cm}

\begin{abstract} Quantum error-correcting codes are studied from classical matrix product
codes point of view. Two methods to construct quantum codes from matrix product codes are
provided. These constructions are applied to obtain numerous new quantum codes, some of them
have better parameters than current quantum codes available.
\end{abstract}

%\vspace*{1cm}

\bf Key Words\rm :  Quantum codes, Matrix product codes,  Hermitian construction

\section{Introduction}
Quantum error-correcting codes play an important role in  quantum communications and quantum computations. After the pioneering  work in \cite{GHR,ZG15,CRSS98}, the theory of quantum codes has developed rapidly in recent decade years. As we know, the approach of constructing new quantum codes which have good parameters is an interesting research field. However, obtaining the parameters of the new quantum codes, especially the  new good  quantum codes, is a difficult problem.  Recently, a lot of new quantum codes have  been constructed by classical linear codes with Hermitian dual containing, which  can be found in \cite{AKS07,AK01,JLLX10,JX14,CLZ15,KZ13,KZ14,G11}).

Matrix product codes  over  finite fields were introduced in \cite{BN01}. Many well-known constructions can be formulated as matrix-product codes. For example, the  $(u|u + v)$-construction, the $(u+v+w|2u+v|u)$-construction, the $(a + x|b + x|a + b + x)$-construction, and etc. The constructed codes mentioned above can be viewed special cases of matrix product codes (See \cite{BN01}). Recently, Galindo et al. in \cite{GHR} constructed
some new quantum codes from matrix-product codes and the Euclidean construction. In \cite{ZG15}, by using generalized Reed-Solomon codes and special matrix with order $2$, Zhang and Ge gave three new classes of quantum MDS codes from generalized Reed-Solomon codes and presented a new construction of quantum codes via matrix-product codes and the Hermitian construction.
Following this  line,  more new quantum codes can be obtained from matrix-product codes and the Hermitian construction. On the one hand, we study a class of matrices over finite fields. By using these matrices, we give a new construction  of quantum codes via matrix-product codes and the Hermitian construction. And our results generalize some previous works in \cite{GHR,ZG15}.
On the other hand, several classes of new  quantum MDS codes are obtained from matrix-product codes and the Hermitian dual containing. Moreover, some new quantum codes obtained in this paper have better
parameters than the quantum codes listed in table online \cite{Edel}.

This paper is organized as follows. Section $2$ recalls the basics about linear codes, matrix-product codes and quantum
codes. In Section $3$,  we give two new constructions of quantum codes by using matrix-product codes. Section $4$, a brief summary of our work is described.
\section{Preliminaries}
The following notations are fixed throughout this paper:
\begin{itemize}
  \item $m,n$ are   positive  integers, and $q$ is a power of prime.
  \item $p$ is  prime, and $\mathbb{F} _{q}$ be the finite field with $q$   elements. $\mathbb{F}_{q}^*=\mathbb{F}_{q}\setminus\{0\}.$
    \item ${\rm Tr}_1^m(\cdot)$  denotes the trace function from $\mathbb{F}_{p^m}$ to $\mathbb{F}_p$, i.e. ${\rm Tr}_1^m(x)=\sum_{i=0}^{m-1}x^{p^i},$ $x\in \mathbb{F}_{p^m}$.
   \item   $\omega=e^{\frac{2\pi\sqrt{ -1}}{p}}$ is a complex primitive $p$-th root of unity.
  \item $\mathbb{F}_{q}^{n}$ denotes the vector space of all $n$-tuples over $\mathbb{F} _{q}$.
\end{itemize}

For any two vectors $\mathbf{a}=(a_1,a_2,\ldots,a_n)\in \mathbb{F}_{q}^{n}$, $\mathbf{b}=(b_1,b_2,\ldots,b_n)\in \mathbb{F}_{q}^{n}$,
the  Euclidean inner product of $\mathbf{a}, \mathbf{b}$ is defined as $$\langle\mathbf{a},\mathbf{b}\rangle_e=\sum_{i=1}^{n}a_{i}b_{i}\in \mathbb{F}_{q} .$$
Let $\mathcal{C}\subseteq \mathbb{F}_{q}^{n}$ be a code of length $n$ over $\mathbb{F}_{q}$, and the Euclidean dual code of $\mathcal{C}$  is defined as
$$C^{\perp_e}=\{\mathbf{x}\in\mathbb{F} _{q}^{n}\mid\langle\mathbf{x},\mathbf{c}\rangle_e=0~\mathrm{for}~\mathrm{all}~\mathbf{c}\in C\}.$$
If $\mathcal{C}$ is linear, then we have $|\mathcal{C}|\cdot|\mathcal{C}^{\perp_e}|=q^n.$

Let $\mathbf{x}=(x_1,x_2,\cdots,x_n) \in \mathbb{F}_{q}^{n} $ be a vector. Let  $w_H(\mathbf{x})$ denote the Hamming weight of $\mathbf{x}$ and $d_H(\mathbf{x},\mathbf{y})$ denote the Hamming distance of $\mathbf{x},\mathbf{y}$. We let $d_H(\mathcal{C})$ denote the minimum Hamming distance  of the code $\mathcal{C}$.
A   code $\mathcal{C}$ of length $n$ over $\mathbb{F} _{q}$ with the minimum Hamming distance $d_H(\mathcal{C})$ is called an $(n,|\mathcal{C}|,d_H(\mathcal{C}))_{q}$ code. If $\mathcal{C}$ is a linear code, then it called   an  $[n,k,d_H(\mathcal{C})]_{q}$ code, where $k$ is the dimension of   $\mathcal{C}$.

%The Hamming weight of  $c=(c_1,c_2,\cdots,c_n) \in \mathcal{C}$ is defined as $$w_H(c)=|\{i\mid c_i\neq0,~1\leq i\leq n \}   |. $$
%The Hamming distance of two codewords $c_1,c_2 \in \mathcal{C}$  is defined as $$d_H(c_1,c_2) = w_H(c_2-c_2).$$ The minimum Hamming distance  of $\mathcal{C}$ is denoted by
%$$d=min\{d_H(c_1,c_2)\mid   c_1\neq c_2 \in \mathcal{C}\}.$$
Let $l$ be a power of prime  and $q=l^2$. For any two vectors $\mathbf{a}=(a_1,a_2,\ldots,a_n) \in \mathbb{F}_{q}^{n}$, $\mathbf{b}=(b_1,b_2,\ldots,b_n) \in \mathbb{F}_{q}^{n}$,
the  Hermitian inner product of $\mathbf{a}, \mathbf{b}$ is defined as $$\langle\mathbf{a},\mathbf{b}\rangle_h=\sum_{i=1}^{n}a_{i}b_{i}^l\in \mathbb{F}_{q} .$$
Let $\mathcal{C}\subseteq \mathbb{F}_{q}^{n}$ be a code of length $n$ over $\mathbb{F}_{q}$, and the Hermitian dual code of $\mathcal{C}$  is defined as
$$\mathcal{C}^{\perp_h}=\{\mathbf{x}\in\mathbb{F} _{q}^{n}\mid\langle\mathbf{x},\mathbf{c}\rangle_h=0~\mathrm{for}~\mathrm{all}~\mathbf{c}\in \mathcal{C}\}.$$
If $\mathcal{C}$ is linear, then we also have $|\mathcal{C}|\cdot|\mathcal{C}^{\perp_{h}}|=q^n.$(See \cite{Fan})
Moreover, it is easy to check that $(\mathcal{C}^{\perp_{h}})^{\perp_{h}}=\mathcal{C}$.

A  linear code $C$ is called Hermitian (Euclidean) dual-containing if $\mathcal{C}^{\perp_{h}}\subseteq \mathcal{C} $  $(\mathcal{C}^{\perp_e}\subseteq \mathcal{C}$).
Let $a=(a_1,a_2,\ldots,a_n)\in \mathbb{F}_{q}^n$, we denote $a^{l}=(a_1^{l},a_2^{l},\ldots,a_n^{l}).$  For a  code $\mathcal{C}$ of length $n$ over $\mathbb{F} _{q}^{n}$, we denote $\mathcal{C}^l$ as
$\{a^{l}\mid~\mathrm{for}~\mathrm{all}~a\in \mathcal{C}\}$.  Hence,  we have that  if $\mathcal{C}$ is linear, then  $\mathcal{C}^{\perp_{h}}=(\mathcal{C}^{l})^{\perp_e}$. Therefore,
$\mathcal{C}$ is Hermitian dual-containing if and only if $(\mathcal{C}^{l})^{\perp_e}\subseteq \mathcal{C}$ which is equivalence to  $\mathcal{C}^{\perp_e}\subseteq \mathcal{C}^{l}$.

\subsection{The Matrix Product Codes}
Let $s \leq m$ and $A=(a_{ij})_{s\times m}$ be an $s\times m$ matrix over $\mathbb{F}_q$. let $\mathcal{C}_1,\cdots,\mathcal{C}_s$ be codes of length $n$ over $\mathbb{F}_q$.
The matrix product codes $\mathcal{C}=[\mathcal{C}_1,\ldots,\mathcal{C}_s]A$ is the set of  all matrix product $[\mathbf{c}_1,\cdots,\mathbf{c}_s]A$, where $\mathbf{c}_i\in \mathcal{C}_i$ are $n\times1$ column vectors for $1\leq j\leq s$.     If $\mathcal{C}_1,\ldots,\mathcal{C}_s$ are all linear codes with generator matrices $G_1,\ldots,G_s$, respectively, then we have $[\mathcal{C}_1,\ldots,\mathcal{C}_s]A$ is a linear code generated by the following  matrix
$$G=\begin{pmatrix}a_{11}G_1&a_{12}G_1&\cdots&a_{1m}G_1\\ a_{21}G_2& a_{22}G_2 & \cdots&a_{2m}G_2 \\ \vdots &\vdots&\cdots& \vdots\\a_{s1}G_s & a_{s2}G_s& \cdots & a_{sm}G_s\end{pmatrix}.$$

For any integer $k$ with $1\leq k \leq s$, we denote that the $i$th rows of $A$ generates a linear code of length $m$ over $\mathbb{F}_q$  by $U_{A}(k)$, where $i=1,2,\cdots,k$.
Let $A_t$ be the matrix consisting of the first $t$ rows of  $A=(a_{ij})_{s\times m}$ . For $1\leq j_1 <j_2<\cdots<j_t\leq m$, we let  $A(j_1,j_2,\ldots,j_t)$ be a $t\times t$ matrix  consisting of columns $j_1,j_2,\ldots,j_t$ of $A_t$.
\begin{define}{\rm \cite{BN01} }\label{def:2.1}
Let the notations be given as above. A matrix $A$ is called a full-row-rank(FRR) matrix if  its row vectors are linearly independent. If $A(j_1,j_2,\ldots,j_t)$ is non-singular for any $1\leq t \leq s$ and  $1\leq j_1 <j_2<\cdots,j_t\leq m$, then $A$ is said to be non-singular by columns (NSC).
\end{define}

In the following, we list some useful results on matrix-product codes, which can be found in \cite{BN01}.
\begin{lemma}{\rm \cite{BN01} }\label{lem:Bn1}
Assume the notations are given as above. Let $\mathcal{C}_i$ be an $[n,k_i,d_i]_{q}$ linear code for $1\leq i\leq s$ and $A=(a_{ij})_{s\times m}$ be an FRR matrix. Let $\mathcal{C}=[\mathcal{C}_1,\cdots,\mathcal{C}_s]A$, then $\mathcal{C}$ is an $[nm,\sum_{i=1}^{s}k_i,d(\mathcal{C})]_{q}$ linear code. Moreover, we have
$$d(\mathcal{C})\geq \mathrm{min}\{d_1d(U_A(1)),d_2d(U_A(2)),\ldots,d_sd(U_A(s)\}.$$
\end{lemma}

\begin{lemma}{\rm \cite{BN01} }\label{lem:Bn2}
Assume the notations are given as above. Let $\mathcal{C}_i$ be an $[n,k_i,d_i]_{q}$ linear code for $1\leq i\leq s$ and $A$ be an $s\times m$ NSC matrix. Let $\mathcal{C}=[\mathcal{C}_1,\cdots,\mathcal{C}_s]A$, then
\begin{itemize}
  \item [{\rm (i)}]  $d(\mathcal{C})\geq d^{*}=\mathrm{min}\{md_1,(m-1)d_2,\ldots,(m-s+1)d_s\}$;

  \item [{\rm (ii)}] If $A$ is also upper-triangular then $d(\mathcal{C})= d^{*}$.
\end{itemize}
\end{lemma}

\begin{lemma}{\rm \cite{BN01} }\label{lem:Bn3}
Let the notations be given as above. Let  $A$ be an  $s\times s$ non-singular matrix and $\mathcal{C}_1,\cdots,\mathcal{C}_s$ be linear codes over $\mathbb{F}_q$, then
$$([\mathcal{C}_1,\ldots,\mathcal{C}_s]A)^{\perp_e}=[\mathcal{C}_1^{\perp_e},\ldots,\mathcal{C}_s^{\perp_e}](A^{-1})^{T}.$$
Furthermore,
\begin{itemize}
  \item [{\rm (i)}] If $A$ is an $s\times s$ NSC matrix, then $$d(\mathcal{C}^{\perp_e})\geq (d^{\perp_e})^{*}=\mathrm{min}\{sd_s^{\perp_e},(s-1)d_{s-1}^{\perp_e},\ldots,d_1^{\perp_e}\};$$
  \item [{\rm (ii)}] If $A$ is an upper-triangular matrix, then $$d(\mathcal{C}^{\perp_e})= (d^{\perp_e})^{*}.$$
\end{itemize}
\end{lemma}

\subsection{Quantum Codes}
Let $q=l^2$ and $l=p^{m}$. Let $V_n=\underbrace{\mathbb{C}^{l^n}=\mathbb{C}^{l}\otimes\cdots\otimes\mathbb{C}^{l}}_n$ be the Hilbert space and let $|x\rangle$ be the vectors of an orthogonal basis of $\mathbb{C}^{l^n}$, where $x\in \mathbb{F}_{l}$. Then $V_n$ has the following orthogonal basis $$\{|c\rangle=|c_1c_2\cdots c_n=|c_1\rangle \otimes|c_2\rangle \otimes\cdots\otimes |c_n\rangle:~c=(c_1,c_2,\ldots,c_n) \in \mathbb{F}_{l}^{n}\}.$$
For $a,b\in \mathbb{F}_{l}$, the unitary linear operators $X(a)$ and $Z(b)$ in $\mathbb{C}^{l}$ are defined as $$X(a)|x\rangle=|x+a\rangle, ~~Z(b)|x\rangle=\omega^{{\rm Tr}_1^m(bx)}|x\rangle.$$

For $\mathbf{a}=(a_1,\ldots,a_n)\in \mathbb{F}_{l}$, we let $X(\mathbf{a})=X(a_1)\otimes\cdots\otimes X(a_n)$ and $Z(\mathbf{a})=Z(a_1)\otimes\cdots\otimes Z(a_n)$ be the tensor products of $n$ error operators. Then $E_n=\{X(\mathbf{a})Z(\mathbf{b}): \mathbf{a},\mathbf{b}\in \mathbb{F}_{l}^n\}$ is an error basis on the complex vector space $\mathbb{C}^{l^n}$, and  $G_n=\{w^cX(\mathbf{a})Z(\mathbf{b}): \mathbf{a},\mathbf{b}\in \mathbb{F}_{l}^n,c\in\mathbb{F}_{p} \}$ is the error group associated with $E_n$.

\begin{define}\label{def:2.5}
Let   quantum code $Q$ of length $n$ be a  subspace  of $V_n$ with dimension $K>1$.  If  $K>2$ and $Q$ detects $d-1$ quantum digits of errors for $d\geq1$, we call  $Q$  to be a symmetric quantum code (SQC), and  denote it by  $((n,K,d))_l$ or $[[n,k,d]]_l$, where $k=log_lK$. Namely, if for every orthogonal pair $|u\rangle,|v\rangle$ in $Q$ with $<u|v>=0$ and every $e\in G_n$ with $W_Q(e)\leq d-1$, $|u\rangle$ and $e|v\rangle$ are orthogonal, i.e.,$<u|e|v>=0$. Such a quantum code is called pure if $<u|e|v>=0$ for any  $|u\rangle$ and $|v\rangle$  in $Q$ and any $e\in G_n$ with $1\leq W_Q(e)\leq d-1$. A quantum code $Q$ with $K=1$ is always pure.
\end{define}
Let us recall the SQC $Q$ construction:
\begin{theorem}{\rm \cite{JX14}}
Let $C$  be  a classical linear $[n,k,d]_{l^2}$ code. If  $C^{\perp_h}\subseteq C$, then there exists an SQC  $Q$  with parameters $[[n,2k-n,\geq d]]_{l}$ that is pure to $d$.
\end{theorem}
To see that an SQC  $Q$ is good in terms of its parameters, we have to introduce the quantum Singleton bound (See \cite{JX14}).
\begin{theorem}{\rm \cite{JX14}}
Let $Q$ be an SQC with parameters $[[n,k,d]]_{l}$. Then $2d\leq n-k+2$.
\end{theorem}

If  an  SQC  $Q$ with parameters $[[n,k,d]]_{l}$ attains the quantum Singleton bound $2d= n-k+2$, then it is called an SQC maximum-distance-separable code (SQCMDS).

\section{New Quantum Codes From Matrix Product Codes}

Throughout this  section, let $l$ be a power of  odd prime and $A =(a_{ij} )$ be an $s\times s$ matrix over $\mathbb{F}_{l^{2}}$. Let  $A^{(l)} = (a_{ij}^{l})$, we have the following result.

\begin{theorem}\label{th:3.1}
Let  $C_i\subset\mathbb{F}_{l^{2}}^{n}$ be an $[n,k_i^{*},d_i]_{l^2}$  code and $C_{i}^{\perp_{h}}\subset C_i$, where $1\leq i\leq s$.  Let $C=[C_1,C_2,\ldots,C_s]A$. If $A^{(l)}A^{T}$  is a diagonal square matrix over $\mathbb{F}_{l^{2}}^{*}$, then $([C_1,C_2,\ldots,C_s]A)^{\perp_{h}}\subset[C_1,~C_2,~\ldots,~C_s]A$, i.e., $C^{\perp_{h}}\subset C$. In particular, if $[(A^{(l)})^{-1}]^{T}=aA$, where $a\in\mathbb{F}_{l^{2}}^{*}$, then $([C_1,C_2,\ldots,C_s]A)^{\perp_{h}}\subset[C_1,~C_2,~\ldots,~C_s]A$.
\end{theorem}
\pf Assume that $A^{(l)}A^{T}=D=\begin{pmatrix}u_{1}&0&\cdots&0\\ 0& u_{2} & \cdots&0\\ \vdots &\vdots&\cdots& \vdots\\0 & 0& \cdots & u_{s}\end{pmatrix}$, with $u_i\in \mathbb{F}_{l^{2}}^*$. Then $A^{(l)}A^{T}D^{-1}=I$. By Lemma~\ref{lem:Bn3} and note that $\mathcal{C}^{\perp_h}=(\mathcal{C}^l)^{\perp_e}$, we have
\begin{eqnarray*}
  ([C_1,\ldots,C_s]A)^{\perp_{h}} &=& ([C_1^{l},\ldots,C_s^{l}]A^{(l)})^{\perp_{e}} \\
   &=& [(C_1^{l})^{\perp_{e}},\ldots,(C_s^{l})^{\perp_{e}}][(A^{(l)})^{-1}]^{T} \\
   &=& [C_1^{\perp_{h}},\ldots,C_s^{\perp_{h}}][(A^{(l)})^{-1}]^{T} \\
   &=& [C_1^{\perp_{h}},\ldots,C_s^{\perp_{h}}][(A^{T})D^{-1}]^{T}\\
   &=& [C_1^{\perp_{h}},\ldots,C_s^{\perp_{h}}]D^{-1}A.
\end{eqnarray*}
Since $D^{-1}=\begin{pmatrix}u_{1}^{-1}&0&\cdots&0\\ 0& u_{2}^{-1} & \cdots&0\\ \vdots &\vdots&\cdots& \vdots\\0 & 0& \cdots & u_{s}^{-1}\end{pmatrix}$ and $u_k^{-1}C_{k}^{\perp_{h}}=C_{k}^{\perp_{h}}$  for $k=1,\ldots,s$, we obtain
$$[C_1^{\perp_{h}},\ldots,C_s^{\perp_{h}}]D^{-1}=[u_{1}^{-1}C_1^{\perp_{h}},\ldots,u_{s}^{-1}C_s^{\perp_{h}}]=[C_1^{\perp_{h}},\ldots,C_s^{\perp_{h}}].$$
Therefore,
$$C^{\perp_{h}}=[C_1^{\perp_{h}},\ldots,C_s^{\perp_{h}}]D^{-1}A=[C_1^{\perp_{h}},\ldots,C_s^{\perp_{h}}]A\subset[C_1,\ldots,C_s]A=C.$$
\qed

%In the following, we construct a class of matrix $A=(a_{ij})_{s\times s}$ whose satisfy  $[(A^{(l)})^{-1}]^{T}=sA$.

Let $s$ be a positive integer with  $s\mid(l^{2}-1)$. Let $G_2=\langle a:a^2=1\rangle$ and $G=\underbrace{G_{2}\times G_2\times\cdots\times G_2}_{r}$. It is easy to check $G$ is an abelian group. Let $\widehat{G}$ be the set of characters of $G$ with respect to $\mathbb{F}_{l^{2}}$. Then we have $\widehat{G}=\{\chi_0,\chi_1,\ldots,\chi_{s-1}\}$, where $\chi_0,\chi_1,\ldots,\chi_{s-1}$ are all  irreducible characters of $G$. For any $j=0,1,\ldots,s-1$, we have  $\chi_{j}(g)^{2}=1$, where $g\in G$. Therefore, we have the following character table
$$ A=\begin{pmatrix}\chi_{0}(g_0)&\chi_{1}(g_0)&\cdots&\chi_{s-1}(g_0)\\\chi_{0}(g_1)&\chi_{1}(g_1)&\cdots&\chi_{s-1}(g_1)\\ \vdots &\vdots&\cdots& \vdots\\\chi_{0}(g_{s-1})&\chi_{1}(g_{s-1})&\cdots&\chi_{s-1}(g_{s-1})\end{pmatrix},$$
where $g_0,g_1,\ldots,g_{s-1}\in G$.  Since $l$ is odd, it follows that $l^{2}=2t+1$ for some integer $t$. Note that
\begin{eqnarray*}
  A^{(l)} &=& \begin{pmatrix}(\chi_{0}(g_0))^{l}&(\chi_{1}(g_0))^{l}&\cdots&(\chi_{s-1}(g_0))^{l}\\(\chi_{0}(g_1))^{l}&(\chi_{1}(g_1))^{l}&\cdots&(\chi_{s-1}(g_1))^{l}\\ \vdots &\vdots&\cdots& \vdots\\(\chi_{0}(g_{s-1}))^{l}&(\chi_{1}(g_{s-1}))^{l}&\cdots&(\chi_{s-1}(g_{s-1}))^{l}\end{pmatrix} \\
  &=& \begin{pmatrix}(\chi_{0}(g_0))^{2t+1}&(\chi_{1}(g_0))^{2t+1}&\cdots&(\chi_{s-1}(g_0))^{2t+1}\\(\chi_{0}(g_1))^{2t+1}&(\chi_{1}(g_1))^{2t+1}&\cdots&(\chi_{s-1}(g_1))^{2t+1}\\ \vdots &\vdots&\cdots& \vdots\\(\chi_{0}(g_{s-1}))^{2t+1}&(\chi_{1}(g_{s-1}))^{2t+1}&\cdots&(\chi_{s-1}(g_{s-1}))^{2t+1}\end{pmatrix} \\
  &=& \begin{pmatrix}\chi_{0}(g_0)&\chi_{1}(g_0)&\cdots&\chi_{s-1}(g_0)\\\chi_{0}(g_1)&\chi_{1}(g_1)&\cdots&\chi_{s-1}(g_1)\\ \vdots &\vdots&\cdots& \vdots\\\chi_{0}(g_{s-1})&\chi_{1}(g_{s-1})&\cdots&\chi_{s-1}(g_{s-1})\end{pmatrix}=A.
\end{eqnarray*}
Since $A$ is invertible, we get
$$ (A^{(l)})^{-1}=A^{-1}=\frac{1}{s}\begin{pmatrix}\chi_{0}(g_0)^{-1}&\chi_{0}(g_1)^{-1}&\cdots&\chi_{0}(g_{s-1})^{-1}\\\chi_{1}(g_0)^{-1}&\chi_{1}(g_1)^{-1}&\cdots&\chi_{1}(g_{s-1})^{-1}\\ \vdots &\vdots&\cdots& \vdots\\\chi_{s-1}(g_{0})^{-1}&\chi_{s-1}(g_{1})^{-1}&\cdots&\chi_{s-1}(g_{s-1})^{-1}\end{pmatrix}.$$
Note that for any $j=0,1,\ldots,s-1$, $\chi_{j}(g)^{2}=1$, we always have $\chi_{j}(g)^{-1}=\chi_{j}(g)$, which implies that $[(A^{(l)})^{-1}]^{T}=\frac{1}{s}A$.
\qed

We need the following corollary to construct quantum codes.

\begin{cor}\label{cor:3.2}
Assume the notations  are given as above. Let $\mathcal{C}_i$  be  an $[n,k_i ,d_i ]_{l^{2}}$ linear codes satisfying $\mathcal{C}_{j} ^{\perp_{h}}\subset \mathcal{C}_{j}$, where $ j = 1,2,3,4$. Then there exists a Hermitian dual containing code over $\mathbb{F}_{l^{2}}$ with the parameter $[4n,k_1 + k_2+k_3+k_4, \geq min \{4d_{1},2d_{2},2d_{3},d_{_4}\}]_{l^{2}}$.
\end{cor}
\pf  Let $G=G_{2}\times G_{2}=\{(1,1),(x,1)(1,y),(x,y):x^2=y^2=1\}$. Then the character table of $G$ is
$$ A=\begin{pmatrix}1&1&1&1\\1&1&-1&-1\\ 1&-1&1&-1\\1&-1&-1&1\end{pmatrix}.$$
By above discussion of $A$  and Theorem~\ref{th:3.1}, we get  the matrix product code $\mathcal{C}=[\mathcal{C}_{1},\mathcal{C}_{2},\mathcal{C}_{3},\mathcal{C}_{4}]A$ satisfies $C^{\perp_{h}}\subset C$. By Lemma~\ref{lem:Bn1}, $\mathcal{C}$  has parameter $[4n,k_1 + k_2+k_3+k_4, \geq min \{4d_{1},2d_{2},2d_{3},d_{_4}\}]_{l^{2}}$.
\qed

The followings can be found in \cite{JLLX10,GHR}.

\begin{lemma}\label{lem:3.3}
Assume the notations  are given as above, then
\begin{itemize}
  \item [{\rm (i)}] there exists a Hermitian dual containing $[l^{2}-1,l^2-d,d]_{l^{2}}$ code for $1\leq d\leq l+1;$
  \item [{\rm (ii)}] there exists a Hermitian dual containing $[l^{2},l^2+1-d,d]_{l^{2}}$ code for $2\leq d\leq l.$
\end{itemize}
\end{lemma}
\begin{lemma}
Let the notations be given as above, then there exists a Hermitian dual containing $[l^{2}+1,l^2+2-d,d]_{l^{2}}$ MDS code for $2\leq d\leq l+1$.
\end{lemma}

Now, we present an approach to construct some  quantum codes.

\begin{theorem}\label{th:3.5}
Let the notations be given as above, then
\begin{itemize}
  \item [{\rm (i)}] there exists an   $[[4l^{2}-4,4l^2+4-4d-\frac{d}{2},\geq d]]_{l}$ quantum code, where $4\leq d\leq l$ and $d\equiv 0~(\mathrm{mod}~4)$.
  \item [{\rm (ii)}] there exists an  $[[4l^{2}-4,4l^2+6-4d-\frac{d+1}{2},\geq d]]_{l}$ quantum code, where $4\leq d\leq l$ and $d\equiv -1~(\mathrm{mod}~4)$.
  \item [{\rm (iii)}] there exists an   $[[4l^{2},4l^2+8-4d-\frac{d}{2},\geq d]]_{l}$ quantum code, where $4\leq d\leq l$ and $d\equiv 0~(\mathrm{mod}~4)$.
  \item [{\rm (iv)}] there exists an  $[[4l^{2},4l^2+6-4d-\frac{d+1}{2},\geq d]]_{l}$ quantum code, where $4\leq d\leq l$ and $d\equiv-1~(\mathrm{mod}~4)$.
  \item [{\rm (v)}] there exists an   $[[4l^{2}+4,4l^2+12-4d-\frac{d}{2},\geq d]]_{l}$ quantum code, where $4\leq d\leq l+1$ and $d\equiv 0~(\mathrm{mod}~4)$.
  \item [{\rm (vi)}] there exists an   $[[4l^{2}+4,4l^2+10-4d-\frac{d+1}{2},\geq d]]_{l}$ quantum code, where $4\leq d\leq l+1$ and $d\equiv -1~(\mathrm{mod}~4)$.
\end{itemize}
\end{theorem}
\pf (i) Let $4\leq d\leq l$ and $d\equiv 0~(\mathrm{mod}~4)$. Let $C_1$ be a Hermitian dual containing code with parameter $[l^{2}-1,l^2-\frac{d}{4},\frac{d}{4}]_{l^{2}}$, and $C_2=C_3$ be  Hermitian dual containing codes with same parameter $[l^{2}-1,l^2-\frac{d}{2},\frac{d}{2}]_{l^{2}}$. Let $C_4$ be a Hermitian dual containing $[l^{2}-1,l^2-d,d]_{l^{2}}$ code.  By Corollary~\ref{cor:3.2},  there exists an $[[4l^{2}-4,4l^2+4-4d-\frac{d}{2},\geq d]]_{l}$ quantum code.

(ii) Let $4\leq d\leq l$ and $d\equiv-1~(\mathrm{mod}~4)$. By  Lemma~\ref{lem:3.3} (ii), taking   $C_1$ to be a Hermitian dual containing $[l^{2}-1,l^2-\frac{d+1}{4},\frac{d+1}{4}]_{l^{2}}$ code, and $C_2=C_3$ are  two Hermitian dual containing $[l^{2}-1,l^2-\frac{d+1}{2},\frac{d+1}{2}]_{l^{2}}$ codes, and $C_4$ is a Hermitian dual containing $[l^{2}-1,l^2-d,d]_{l^{2}}$ code, we have that there exists an $[[4l^{2}-4,4l^2+6-4d-\frac{d+1}{2},\geq d]]_{l}$ quantum code by Corollary 3.2.

Other cases are proven similarly.
\qed

\begin{remark} By Theorem~\ref{th:3.5}, we obtain some new quantum codes.  Comparing to the quantum codes obtained in \cite{Edel},   new quantum codes in Table~$1$ have  better parameters.
\end{remark}
\begin{table}
\caption{QUANTUM CODES COMPARISON}
\begin{center}\begin{tabular}{|c|c|}
\hline
new quantum codes & quantum codes from [13]  \\
\hline
$[[96,86,\geq 4]]_5$ & $[[96,82,4]]_5$  \\
\hline
$[[104,94,\geq 4]]_5$ & $[[104,94,4]]_5$  \\
\hline
$[[192,182,\geq 4]]_7$ & $[[192,182,3]]_7$\\
\hline
$[[192,170,\geq 7]]_7$ & $[[192,170,5]]_7$ \\
\hline
$[[200,190,\geq 4]]_7$ & $[[200,188,4]]_7$\\
\hline
$[[200,172,\geq 8]]_7$ & $[[200,172,8]]_7$\\
\hline
$[[320,310,\geq 4]]_9$ &  $[[320,310,3]]_9$ \\
\hline
$[[320,292,\geq 8]]_9$ & $[[320,284,8]]_9$ \\
\hline
$[[320,298,\geq 7]]_9$ &  $[[320,298,5]]_9$ \\
\hline
$[[328,318,\geq 4]]_9$ & $[[328,318,4]]_9$ \\
\hline
\end{tabular}\end{center}
\end{table}
\begin{lemma}\label{th:main1}
Let  $\mathcal{C}_i$ be an $[n,k_i,d_i]_{l^{2}}$ linear code and $\mathcal{C}_{i}^{\perp_{h}}\subset \mathcal{C}_i$  for $1\leq i\leq s$.  If $\mathcal{C}_1\subset \mathcal{C}_2\subset\cdots\subset \mathcal{C}_s$ and $A$  is an $s\times s$ NSC upper-triangular matrix, then
$$([\mathcal{C}_1,\mathcal{C}_2,\ldots,\mathcal{C}_s]A)^{\perp_{h}}\\\subset[\mathcal{C}_1,\mathcal{C}_2,\ldots,\mathcal{C}_s]A.$$
\end{lemma}
\pf Since $A$ is an $s\times s$ NSC  upper-triangular matrix, and $[(A^{(l)})^{-1}]^{T}$ is an $s\times s$ lower-triangular matrix, they are of the forms:
$$ A=\begin{pmatrix}a_{11}&a_{12}&\cdots&a_{1s}\\0& a_{22} & \cdots&a_{2s }\\ \vdots &\vdots&\cdots& \vdots\\0 & 0& \cdots & a_{ss}\end{pmatrix},$$
and
$$[(A^{(q)})^{-1}]^{T}=\begin{pmatrix}b_{11}&0&\cdots&0\\ b_{21}& b_{22} & \cdots&0\\ \vdots &\vdots&\cdots& \vdots\\b_{s1} & b_{s2 }& \cdots & b_{ss}\end{pmatrix},$$
where $a_{11}a_{22}\cdots a_{ss}\neq0$. Obviously, we have
$$[\mathcal{C}_1,\mathcal{C}_2,\cdots,\mathcal{C}_s]A=\{(a_{11}c_1,a_{12}c_1+a_{22}c_2,\cdots,a_{1s}c_1+a_{2s}c_2+\cdots+a_{ss}c_s)\mid c_i\in \mathcal{C}_i, 1\leq i \leq s\}$$
and
\begin{eqnarray*}
    & & [C_1^{\perp_{h}},C_2^{\perp_{h}},\ldots,C_s^{\perp_{h}}][(A^{(l)})^{-1}]^{T} \\
    &=& \{b_{11}c_1^{\perp_{h}}+b_{21}c_2^{\perp_{h}}+\cdots+b_{s1}c_s^{\perp_{h}},b_{22}c_2^{\perp_{h}}+\cdots+b_{s2}c_s^{\perp_{h}},\cdots,b_{ss}c_s^{\perp_{h}}\mid \\
    & &c_1^{\perp_{h}}\in C_1^{\perp_{h}},c_2^{\perp_{h}}\in C_2^{\perp_{h}},\cdots,c_s^{\perp_{h}}\in C_s^{\perp_{h}}\}.
\end{eqnarray*}

Now, we prove  for any $(b_{11}c_1^{\perp_{h}}+b_{21}c_2^{\perp_{h}}+\cdots+b_{s1}c_s^{\perp_{h}},b_{22}c_2^{\perp_{h}}+\cdots+b_{s2}c_s^{\perp_{h}},\cdots,b_{ss}c_s^{\perp_{h}})
\in[C_1^{\perp_{h}},C_2^{\perp_{h}},\cdots,C_s^{\perp_{h}}][(A^{(l)})^{-1}]^{T},$ there exist $\widetilde{c}_1\in C_1,\widetilde{c}_2\in C_2,\cdots,\widetilde{c}_s\in C_s$ such that
$$[\widetilde{c}_1,\widetilde{c}_2,\ldots,\widetilde{c}_s]A=(b_{11}c_1^{\perp_{h}}+
b_{21}c_2^{\perp_{h}}+\cdots+b_{s1}c_s^{\perp_{h}},b_{22}c_2^{\perp_{h}}+
\cdots+b_{s2}c_s^{\perp_{h}},\cdots,b_{ss}c_s^{\perp_{h}})$$
or
\begin{eqnarray*}
    &&(a_{11}\widetilde{c}_1,a_{12}\widetilde{c}_1+a_{22}\widetilde{c}_2,\cdots,a_{1s}\widetilde{c}_1+a_{2s}\widetilde{c}_2+\cdots+a_{ss}\widetilde{c}_s) \\
   &=& (b_{11}c_1^{\perp_{h}}+b_{21}c_2^{\perp_{h}}+\cdots+b_{s1}c_s^{\perp_{h}}, b_{22}c_2^{\perp}+\cdots+b_{s2}c_s^{\perp},\cdots,b_{ss}c_s^{\perp}).
\end{eqnarray*}
Let $\beta_1=b_{11}c_1^{\perp_{h}}+b_{21}c_2^{\perp_{h}}+\cdots+b_{s1}c_s^{\perp_{h}}$, $\beta_2=b_{22}c_2^{\perp_{h}}+\cdots+b_{s2}c_s^{\perp_{h}}, \cdots,\beta_s=b_{ss}c_s^{\perp_{h}}.$
For $k=1$, we let $\widetilde{c}_1=\frac{1}{a_{11}}\beta_1$, then obtain $\widetilde{c}_1\in C_1$ and $a_{11}\widetilde{c}_1=\beta_1$.
For $k=2,3,\cdots,s$, let $$\widetilde{c}_k=-\frac{1}{a_{kk}}(a_{1k}\widetilde{c}_1+a_{2k}\widetilde{c}_2+\cdots+a_{k-1,k}\widetilde{c}_{k-1}-\beta_k),$$ we have  $\widetilde{c}_k\in C_k$  and $a_{1k}\widetilde{c}_1+a_{2k}\widetilde{c}_2+\cdots+a_{k-1,k}\widetilde{c}_{k-1}+a_{kk}\widetilde{c}_{k}=\beta_k$,
as desired.
\qed

\begin{lemma}{\rm \cite{KZ14}}\label{lem:kai1}
Assume the notations are given as above. Let $l\equiv1~(\mathrm{mod}~4)$ and $n=l^2+1$,  suppose $t=\frac{n}{2}$. If $C$ is a negacyclic code of length $n$ with defining set $Z=\bigcup_{i=0}^{\delta}C_{t-2i}$, where $0\leq\delta\leq\frac{l-1}{2}$, then $C^{\perp_{h}}\subset C$.
\end{lemma}
\begin{lemma}{\rm \cite{KZ14}}\label{lem:kai2}
 Let $n=\frac{l^2+1}{2}$,where $l$ is a power of an odd prime. If $C$ is a negacyclic code of length $n$ with defining set $Z=\bigcup_{i=0}^{\delta}C_{2i-1}$, where $1\leq\delta\leq\frac{l-1}{2}$, then $C^{\perp_{h}}\subset C$.
\end{lemma}
\begin{theorem}\label{th:main2}
Assume the notations are given as above. Let $l\equiv1~(\mathrm{mod} ~4)$ and $n=l^2+1$.  Let $t=\frac{n}{2}$, and let $C_j$ be a negacyclic code of length $n$ with defining set $Z_j=\bigcup_{i=0}^{\delta_{j}}C_{t-2i}$ for $j=1,2,3$, where $1\leq\delta_3\leq\delta_2\leq\delta_1\leq\frac{l-1}{2}$, then there exists an $[[3l^{2}+3,3l^2-4\delta_1-4\delta_2-4\delta_3-3,\geq 2(\delta_3+1)]]_{l}$ quantum code.
\end{theorem}
\pf It is easy to see that $C_1\subset C_2\subset C_3$. By Lemma~\ref{lem:kai1}, we have $C_{1}^{\perp_{h}}\subset C_1,~C_{2}^{\perp_{h}}\subset C_2$, and $C_{3}^{\perp_{h}}\subset C_3$.
Let
 $$A=\begin{pmatrix}1&1&1\\0&2&1\\0&0&1\end{pmatrix},$$
then we have $A$ is an $3\times 3$ NSC upper-triangular matrix. By Lemma 3.7, we have
$$([C_1,C_2,C_3]A)^{\perp_{h}}\\\subset[C_1,C_2,C_3]A.$$
By Lemma~\ref{lem:Bn2}, we get $[C_1,C_2,C_3]A$ is an $[3l^2+3,3l^2+3-\delta_1-\delta_2-\delta_3,\geq d(C_3)]_{l^{2}}$ code. Then by the Hermitian construction, there exists an $[[3l^{2}+3,3l^2+3-2\delta_1-2\delta_2-2\delta_3,\geq d(C_3)]]_{l}$ quantum code.
\qed

Now, we use negacyclic codes to construce  new quantum  codes by Lemma 3.7 and 3.10. We first recall some basic results about negacyclic codes (see \cite{G11}). Since $x^n+1=(x^{2n}-1)/(x^2-1)$ , the roots of  $x^n+1$ are the roots of $x^{2n}-1$ which are not roots of $x^n-1$ in some extension field of $\mathbb{F}_{l^{2}}$. Let $m$ be the multiplicative order of $l^2$ modulo $2n$. Then, $2n|(l^{2m}-1)$. Let $\eta$ be a primitive $2n$th root of unity in $\mathbb{F}_{l^{2m}}$. Then,  the roots of $x^n+1$ are $\eta^{1+2i}, 0\leq i\leq n-1$. Let $C_i$ denote the $l^2$-cyclotomic coset modulo $2n$ containing $i$, and $m_i$  the size of this coset, i.e., $C_i=\{i,il^2,\ldots,il^{2(m_i-1)}\}$.  A $l^2$-ary linear code of length $n$ is negacyclic if $C$ is invariant under the permutation $\tau$ of $\mathbb{F}_{l^{2}}$
$$\tau(c_0,c_1,\ldots,c_{n-1})=(-c_{n-1},c_0,c_1,\ldots,c_{n-2}).$$
Each codeword $c=(c_0,c_1,\ldots,c_{n-1})$ is customarily identified with its polynomial representation $c(x)=c_0+c_1x+\cdots+c_{n-1}x^{n-1}$, and the code $C$ is in turn viewed as the set of all polynomial representations of its codewords. Then, in the quotient ring $\mathbb{F}_{l^{2}}[x]/\langle x^n+1\rangle$, $xc(x)$ corresponds to a negacyclic shift of $c(x)$. This shows that a $l^2$-ary negacyclic code $C$ of length $n$ is precisely an ideal of $\mathbb{F}_{l^{2}}[x]/\langle x^n+1\rangle$. Thus, $C$ can be generated by a monic divisor $g(x)$ of $x^n+1$. Let $\mathcal{O}_{2n}$ be the set of all odd integers from $1$ to $2n$. The defining set of a negacyclic code $C=\langle g(x)\rangle$ of length $n$ is the set $Z=\{i\in \mathcal{O}_{2n}\mid \eta^i\mathrm{ ~is~ a ~root ~of~} g(x)\}$. Obviously, the define set is a union of some $l^2$-cyclotomic cosets modulo $2n$ and $\mathrm{dim}(C)=n-|Z|$.

\begin{example}
By   Theorem~\ref{th:main2}, taking some special values of $l$, we obtain the following new quantum codes.
\begin{center}
\begin{tabular}{c|c|l}
  \hline
  $l$ & $n$ &  {\rm New quantum  codes}    \\\hline
  $5$ & $26$ & $[[78,60,\geq4]]_5$,$[[78,48,\geq6]]_5$   \\\hline
  $9$ & $82$ & $[[246,228,\geq4]]_9$, $[[246,216,\geq6]]_9$, $[[246,204,\geq8]]_9$,  $[[246,192,\geq10]]_9$   \\\hline
  $13$ & $170$ & $[[510,492,\geq4]]_{13}$, $[[510,480,\geq6]]_{13}$, $[[510,468,\geq8]]_{13}$,    \\
    &   & $[[510,456,\geq10]]_{13}$, $[[510,444,\geq12]]_{13}$, $[[510,432,\geq14]]_{13}$   \\\hline
  $17$ & $290$ & $[[870,852,\geq4]]_{17}$, $[[870,840,\geq6]]_{17}$, $[[870,828,\geq8]]_{17}$, $[[870,816,\geq10]]_{17}$,   \\
  &   &  $ [[870,804,\geq12]]_{17}$, $[[870,792,\geq14]]_{17}$, $[[870,780,\geq16]]_{17}$,   $[[870,760,\geq18]]_{17}$.    \\
  \hline
\end{tabular}
\end{center}

\end{example}

%\begin{example}
%Let $l=5$. Then, $n=l^2+1=26$. Applying Theorem~\ref{th:main2}, we obtain two new quantum MDS codes with parameters $[[78,72,4]]_5$ and $[[78,68,6]]_5$.
%\end{example}
%
%\begin{example}
%Let $l=9$. Then, $n=l^2+1=82$. Applying Theorem~\ref{th:main2}, we obtain six new quantum MDS codes with parameters $[[246,240,4]]_9$, $[[246,236,6]]_9$, $[[246,232,8]]_9$, $[[246,228,10]]_9$, $[[246,232,8]]_9$, and $[[246,228,10]]_9$.
%\end{example}
%
%\begin{example}
%Let $l=13$. Then, $n=l^2+1=170$. Applying Theorem~\ref{th:main2}, we obtain six new quantum MDS codes with parameters $[[510,504,4]]_{13}$, $[[510,500,6]]_{13}$, $[[510,496,8]]_{13}$, $[[510,492,10]]_{13}$, $[[510,488,12]]_{13}$, and $[[510,484,14]]_{13}$.
%\end{example}
%
%\begin{example}
%Let $l=17$. Then, $n=l^2+1=290$. Applying Theorem~\ref{th:main2}, we obtain eight new quantum MDS codes with parameters $[[870,864,4]]_{17}$, $[[870,860,6]]_{17}$, $[[870,856,8]]_{17}$, $[[870,852,10]]_{17}$, $[[870,848,12]]_{17}$, $[[870,844,14]]_{17}$, $[[870,840,16]]_{17}$, and $[[870,836,18]]_{17}$.
%\end{example}

By a similar proof as that of Theorem ~\ref{th:main2} and making use of Lemma~\ref{lem:kai2}, we have the following result.

\begin{theorem}\label{th:main3} Let $n=\frac{l^2+1}{2}$,where $l\geq 7$ is a power of an odd prime. If $C_j$ is a negacyclic code of length $n$ with defining set $Z_j=\bigcup_{i=0}^{\delta_{j}}C_{2i-1}$ for $j=1,2,3$, where $1\leq\delta_1<\delta_2<\delta_3\leq\frac{l-1}{2}$, then there exists an $[[3n,3n-2\delta_1-2\delta_2-2\delta_3,\geq d(C_3)]]_{l^{2}}$ quantum code.
\end{theorem}
\begin{example}
By   Theorem~\ref{th:main3}, taking some special values of $l$, we also obtain some new quantum codes.
\begin{center}
\begin{tabular}{c|c|l}
  \hline
  $l$ & $n$ &   {\rm New quantum codes} \\\hline
  $7$ & $25$ & $[[75,63,\geq3]]_7,[[75,51,\geq5]]_7,[[75,39,\geq7]]_7$  \\\hline
  $11$ & $61$ & $[[183,171,\geq3]]_{11},[[183,159,\geq5]]_{11},[[183,147,\geq7]]_{11},[[183,135,\geq9]]_{11},[[183,132,\geq11]]_{11}.$  \\\hline
  $13$ & $85$ & $[[255,243,\geq3]]_{13},[[255,231,\geq5]]_{13},[[255,219,\geq7]]_{13}, [[255,207,\geq9]]_{13},[[255,195,\geq11]]_{13},$  \\
    &   & $ [[255,183,\geq13]]_{13}.$  \\\hline
  $17$ & $145$ & $[[435,423,\geq3]]_{17},[[435,411,\geq5]]_{17},[[435,399,\geq7]]_{17}, [[435,384,\geq9]]_{17},[[435,374,\geq11]]_{17},$   \\
   &  &  $[[435,363,\geq13]]_{17},[[435,351,\geq15]]_{17},[[435,439,\geq17]]_{17}.$  \\
  \hline
\end{tabular}
\end{center}
\end{example}
%\begin{example}
%Taking $l=7$, we have $n=\frac{l^2+1}{2}=25$. By Theorem~\ref{th:main3},  two new quantum MDS codes obtained with with the following parameters$$[[75,67,5]]_7,[[75,63,7]]_7.$$
%\end{example}
%
%\begin{example}
%When $l=11$, then $n=\frac{l^2+1}{2}=61$. By applying Theorem~\ref{th:main3}, we get four new quantum MDS codes with the following parameters $$[[183,175,5]]_{11},[[183,171,7]]_{11},[[183,167,9]]_{11},[[183,163,11]]_{11}.$$
%\end{example}

%\begin{example}
%Let $l=13$, we have $n=\frac{l^2+1}{2}=85$. By using Theorem~\ref{th:main3},    five new quantum MDS codes are obtained with the following parameters $$[[255,247,5]]_{13},[[255,243,7]]_{13}, [[255,239,9]]_{13},[[255,235,11]]_{13}, [[255,231,13]]_{13}.$$
%\end{example}
%
%\begin{example}
%Taking $l=17$, then $n=\frac{l^2+1}{2}=145$. From Theorem~\ref{th:main3}, we get seven new quantum MDS codes with the following parameters $$[[435,427,5]]_{17},[[435,423,7]]_{17}, [[435,419,9]]_{17},[[435,415,11]]_{17}$$  $$[[435,411,13]]_{17},[[435,407,15]]_{17},[[435,403,17]]_{17}.$$
%\end{example}

\section{Conclusion}
We give two   methods to construct  quantum  codes from matrix-product codes.
From our main results, we obtain some good quantum codes and we believe that more good
quantum codes can be obtained from matrix-product codes. \\

\textbf{Acknowledgements}

This work was supported by Research Funds of Hubei Province, Grant
No. D20144401.

\end{document}